# Temperature-dependent excitonic light manipulation with atomically-thin optical elements

Ludovica Guarneri[1], Qitong Li[2], Thomas Bauer[1], Jung-Hwan Song[2], Ashley P. Saunders[3], Fang Liu[3], Mark L. Brongersma[2*], and Jorik van de Groep[1*]

[1] *Van der Waals-Zeeman Institute, Institute of Physics, University of Amsterdam, Amsterdam, 1098 XH, the Netherlands*
[2] *Geballe Laboratory for Advanced Materials, Stanford University, Stanford, CA 94305, USA*
[3] *Department of Chemistry, Stanford University, Stanford, CA 94305, USA*

[*] brongersma@stanford.edu; j.vandegroep@uva.nl

**Abstract**
Monolayer 2D semiconductors, such as $WS_2$, exhibit uniquely strong light-matter interactions due to exciton resonances that enable atomically-thin optical elements. Similar to geometry-dependent plasmon and Mie resonances, these intrinsic material resonances offer coherent and tunable light scattering. Thus far, the impact of the excitons' temporal dynamics on the performance of such excitonic metasurfaces remains unexplored. Here, we show how the excitonic decay rates dictate the focusing efficiency of an atomically-thin lens carved directly out of exfoliated monolayer $WS_2$. By isolating the coherent exciton radiation from the incoherent background in the focus of the lens, we obtain a direct measure of the role of exciton radiation in wavefront shaping. Furthermore, we investigate the influence of exciton-phonon scattering by characterizing the focusing efficiency as a function of temperature, demonstrating an increased optical efficiency at cryogenic temperatures. Our results provide valuable insights in the role of excitonic light scattering in 2D nanophotonic devices.

**Main text**
Optical metasurfaces are dense arrays of resonant nanostructures that collectively manipulate the flow of light to perform an optical function. Careful engineering of the nanostructure design and spatial arrangement affords (near-)arbitrary control over the phase and amplitude of the scattered light with subwavelength spatial resolution, enabling flat optical elements with functionalities beyond conventional bulk optical components[1]. The functionality of optical metasurfaces most commonly relies on light scattering by plasmon or Mie resonances in metallic and dielectric nanostructures, respectively. More recently, excitons in 2D semiconductors such as monolayer transition metal dichalcogenides (TMDs), have emerged as new type of resonance that can be leveraged to realize mutable, flat optics[2,3]. Due to quantum confinement and reduced dielectric screening[4], excitons in monolayer TMDs significantly impact the optical behavior of the material, even at room temperature[5]. The resulting strong and tunable light-matter interaction and atomic thickness offer a new playground for the design of next-generation metasurfaces. The intrinsic nature of exciton resonances in 2D-TMDs renders their spectral properties most notably dependent on the band structure of the material as opposed to the geometry as with plasmon and Mie-resonators. This allows for their facile integration in more complex architectures[6,7]. Furthermore, light scattering by these resonances can be largely and reversibly manipulated via electrostatic free carrier injection[7–9], temperature[7,8], strain[10–12], and external fields[13], enabling actively-tunable nanophotonic devices.

In most of the initial work on 2D semiconductors, a strong excitonic response is obtained in mechanically exfoliated, single-crystal flakes, limited to lateral sizes of a few tens of microns. These high-quality monolayers are typically integrated in multilayered and complex photonic structures that collectively govern the device's optical functionality. At the same time, the strong light-matter interaction within a single TMD monolayer poses an intriguing opportunity to realize atomically-thin optical elements. In the limit of single layer metasurfaces, the optical function is dictated by the nanopatterned monolayer only, without influences of external (van-der-Waals heterostructure) cavities[7,8], or plasmon resonances in nanopatterned electrodes[9,14]. However, to obtain a complex optical functionality such as wavefront shaping by merely nanopatterning the TMD monolayer, the in-plane dimensions of small flakes are typically the limiting factor in achieving the desired functionality. The size obstacle can be resolved using chemical-vapor deposition (CVD) growth techniques. In a recent demonstration[15], we showed how the focusing efficiency of a large-area atomically-thin lens, can be modulated through electrostatic manipulation of the exciton resonance through carrier injection. However, in contrast to small-area exfoliated flakes, the relatively poor quality of the currently used CVD-grown monolayers, penalizes the efficiency of the optical components, hindering the fundamental study of the role of excitonic decay mechanisms in the optical function.

Recently, gold-assisted exfoliation approaches were developed as a key route to obtain high-quality, large area monolayer TMDs[16], opening new possibilities to realize large-scale hybrid metasurfaces[14,17]. Here, we leverage this technique to fabricate a high-quality, large-area and atomically-thin zone-plate lens (diameter = 500 μm) by directly patterning a millimeter-sized exfoliated monolayer of $WS_2$. Using this model system, we explore wavefront shaping at the ultimate limit of a single atomic layer to assess the impact of excitonic decay rates onto the optical efficiency of our lens. To do this, we measure the lens' focusing efficiency and study its spectral line shape. We directly link the focusing efficiency to the optical susceptibility of monolayer $WS_2$, highlighting the intrinsic relation between the excitonic decay rates and the functionality of 2D metasurfaces. To explore the impact of the exciton's quantum yield, we then systematically study the performance of the lens as a function of temperature and observe a significant increase in the optical efficiency in the limit where exciton-phonon scattering is suppressed. Overall, our results show how the optical functionality of the lens is dominantly governed by the interplay of different excitonic decay channels, providing key insight in the design of novel 2D excitonic metasurfaces.

To study the role of excitonic light scattering in large-area atomically-thin optical elements, we employ a $(2 \times 3)$ mm² high-quality monolayer of $WS_2$ directly exfoliated onto a sapphire substrate [16]. Next, using electron-beam lithography and reactive-ion etching, we pattern the monolayer into a zone plate lens with a diameter of 500 μm and nominal 1 mm focal length for $\lambda = 615$ nm wavelength (photon energy = 2.0 eV). Figure 1a shows a bright-field white light optical microscope image of the lens. The bright and darker regions correspond to the $WS_2$-covered and bare sapphire substrate, respectively.

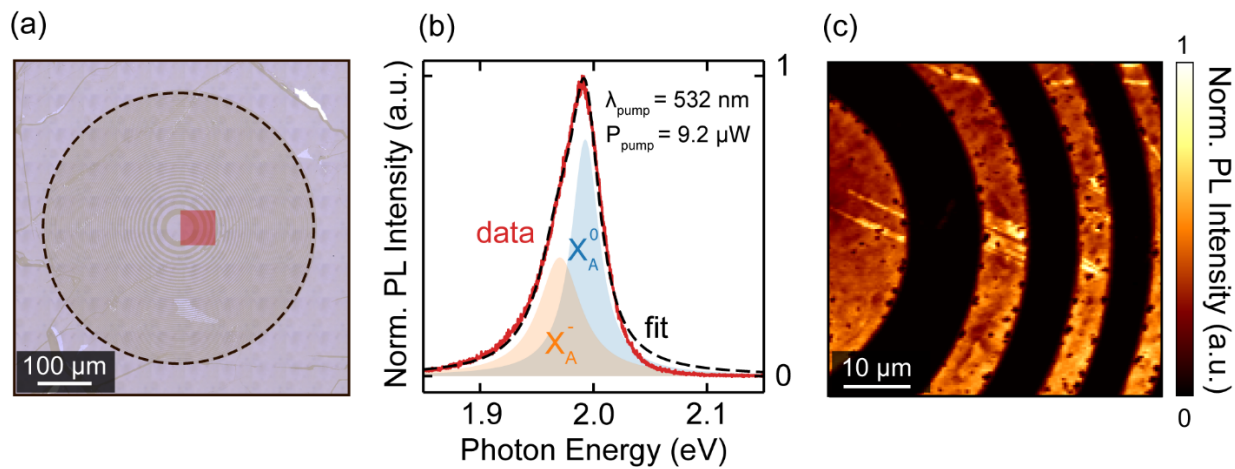

**Figure 1: Large-area atomically-thin lens.** (a) Microscope image of the zone plate lens ($d$ = 500 μm) patterned into a large-area exfoliated monolayer of WS$_2$ on sapphire. The dashed black line outlines the lens contour. Additional information about the sample geometry can be found in Supporting Information section 1. (b) Solid red line: normalized photoluminescence (PL) spectrum of the WS$_2$ monolayer measured in the central area of the lens (one pixel in c). Dashed black line: double-Lorentzian fit to the data. $X_A^0$ (shaded blue) and $X_A^-$ (shaded orange) indicate the relative contribution of the neutral exciton (spectral center at 1.993 eV) and negatively charged trion (spectral center at 1.970 eV) to the PL spectrum, respectively. $\lambda_{pump}$ and $P_{pump}$ are the wavelength and power of the excitation laser used for the PL measurement, respectively. (c) Spatially resolved map of the exciton PL (integrated over the FWHM of $X_A^0$, ~30 meV) for the central area indicated by the red square in a. Additional characterization using Raman mapping spectroscopy further corroborates the presence of monolayer WS$_2$ and its nanopattern (Fig. S1).

We measure the photoluminescence (PL) in the central region of the lens (Fig. 1b) and observe a clear peak around 2 eV which is characteristic of the excitonic emission of monolayer WS$_2$[18–20]. For the neutral excitons ($X_A^0$), we observe an excitonic linewidth of 30 meV at 1.993 eV, circa 10 meV narrower than CVD-grown samples[21] and in line with small-scale exfoliated non-encapsulated monolayers at room temperature[22]. The measured linewidth is indicative of the material quality, emphasizing the benefit of the gold-assisted exfoliation method compared to CVD-grown layers.

We use spatially resolved mapping of the PL signal to verify the successful patterning of the WS$_2$ monolayer into the desired zone plate design (Fig. 1c). The bright regions reveal minor spatial variations in the PL signal in the form of intensity variations and dark spots, which we attribute to residual strain gradients and local imperfections, respectively.

To study the role of exciton-enhanced light scattering in our zone plate lens, we start by characterizing the function of the lens at room temperature. A wavelength-tunable and collimated supercontinuum laser illuminates the sample from the bottom through the transparent sapphire substrate. The lens' focus forms ~1 mm above the sample surface, which we characterize by mapping the light intensity in 3D using a confocal microscope coupled to an avalanche photodiode (Fig. S2).

Figure 2a shows the *x-z* intensity map and corresponding crosscuts for $\lambda$ = 620 nm, which highlight that a clear and intense focus is formed despite the atomically-thin structure of the lens. Sectioning along z allows us to determine the focal height 996.4 μm above the lens surface, close to the designed focal point. To quantify the lens' performance, we map the focal plane in 100 nm steps, which shows the characteristic Airy pattern of a diffraction-limited, focused light-beam (Fig. 2b, see also Supporting Information section 2). We then analyze the intensity in the focus by fitting and spatially integrating the 2D Airy pattern. The background intensity in this image has non-negligible coherent contributions (e.g. directly transmitted light) and incoherent contributions (e.g. reflections from other interfaces in the sample or optics, and dephased exciton radiation). Both can be extracted in the fitting procedure to quantify the true focusing efficiency (see Supporting Information section 2 for details). It is worth noting that reflection measurements from extended flakes are typically used to try and quantify the coherent scattering contribution from excitons. In such measurements it is impossible to extract background contributions, and this may lead to an overestimation of the coherent exciton radiation. In our analysis, we obtain the focusing efficiency spectrum by normalizing the background-free and area-integrated power in the focus by the power incident on the lens surface: i.e. the fraction of the incident power that is redirected into the lens' focal spot (Fig. 2c).

The focusing efficiency spectrum shows three prominent features. First, it reveals an asymmetric Fano-like line shape centered around the main exciton resonance of the $WS_2$ monolayer ($\lambda \sim 620$ nm), which emphasizes that the exciton resonance plays a key role in the lens' function. Second, the overall focusing efficiency is small – below 0.1% absolute – which is a direct result of the atomic-scale optical pathlength in the material. Despite this low efficiency, a clear focus is observed (Fig. 2b), which emphasizes the potential of wave front manipulation with only a single monolayer. Third, the peak value at $\lambda \sim 627$ nm is >3.5 higher than the monotonously decreasing non-resonant (background) efficiency, which highlights the strongly enhanced light-matter interaction offered by coherent exciton radiation[23–25]. These results thus clearly demonstrate that the functionality of atomically-thin and large-area optical elements is strongly affected by the material's exciton resonance and its impact on the optical properties of the monolayer.

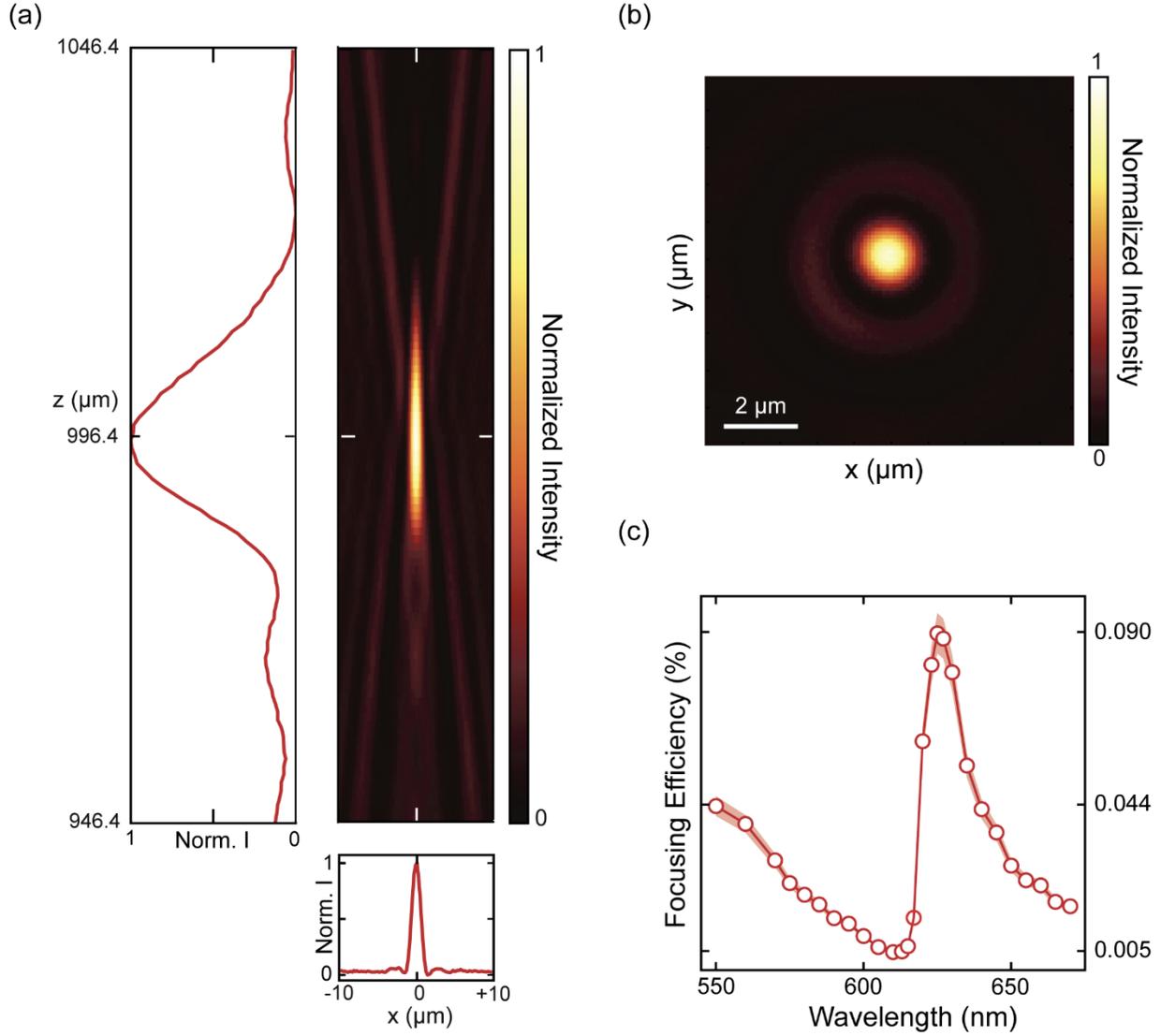

**Figure 2: Room temperature characterization of focal shape and focusing efficiency.** (a) Measured intensity distribution along the *x-z* plane. Cross sections along the optical axis and in the focal plane are also shown. (b) x-y cross section of the intensity distribution at the focal plane showing a 2D Airy pattern. Both a and b are shown for $\lambda = 620$ nm. (c) Room temperature focusing efficiency spectrum centered around the exciton resonance wavelength. The shaded region indicates the measurement error corresponding to two standard deviations (see SI section 3 for error analysis). See SI section 4 for a theoretical comparison to the focusing efficiency of the zone-plate lens.

To further explain the spectral line shape of the focusing efficiency, we quantify the role of the exciton radiation in the optical properties of the monolayer. Fig. 3a shows a reflectance spectrum measured in the center of the zone plate lens. A clear and roughly symmetrically-shaped peak is seen around $\lambda = 620$ nm due to the monolayer's exciton resonance. From this reflectance spectrum, we follow the conventional approach to retrieve the complex susceptibility $\chi(E) = \chi_1(E) + i\chi_2(E)$ of the monolayer $WS_2$, by a Kramers-Kronig constrained analysis[5], with $E$ the photon energy of the incident light. Following the procedure outlined in ref.[5], we then fit this susceptibility (that was retrieved purely numerically) using a physically meaningful model to extract the excitonic decay rates. We use a constant background term and a Lorentzian oscillator[7–9] for each of the ground states of the three main exciton resonances of monolayer WS$_2$ [5]:

$$\chi(E) = \chi_\infty - \sum_{j=A,B,C} \frac{\hbar c}{E_j \cdot t} \cdot \frac{\hbar \gamma_{r_j}}{(E-E_j)+i\frac{\hbar \gamma_{nr_j}}{2}} \qquad (1)$$

Here, $\chi_\infty$ is the constant term to account for higher-energy electronic transitions, $c$ is the speed of light, $E_j$ is the central energy of the $j^{th}$ exciton resonance, and $t$ is the monolayer thickness. $\gamma_{r_j}$ and $\gamma_{nr_j}$ are the radiative and non-radiative decay rates of the $j^{th}$ exciton, respectively. We use $t = 6.18$ Å corresponding to the interlayer spacing of bulk WS$_2$ [5], and from the fit we find $\chi_\infty = 14.23$, $E_A = 1.997\ eV$, $\hbar\gamma_{r_A} = 2.7\ meV$, and $\hbar\gamma_{nr_A} = 40.1\ meV$ for the main exciton resonance. Note that the reported linewidth includes the broadening due to the trion contribution. The complete fitting procedure is detailed in Supporting Information section 5.

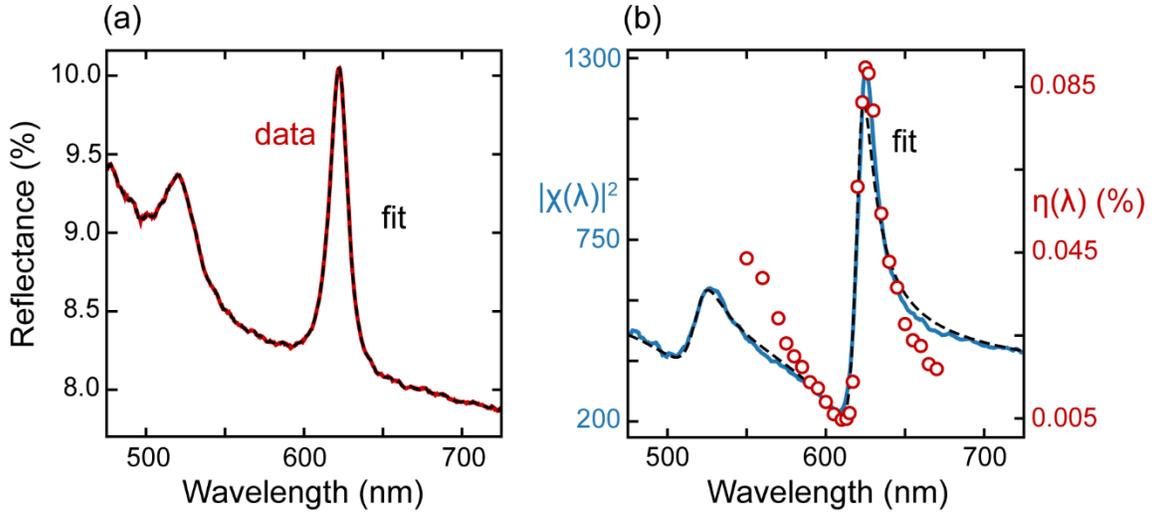

**Figure 3: Optical properties of the monolayer WS$_2$.** (a) Solid red line: reflectance spectrum of monolayer WS$_2$ on sapphire. Dashed black line: numerical fit to the reflectance spectrum using a transfer-matrix model and a multi-Lorentzian oscillator model for the susceptibility. (b) The square modulus of the monolayer's susceptibility (solid blue line), fit to the susceptibility using $\chi_\infty$ and three Lorentzian oscillators for the excitons (dashed black line) and experimental focusing efficiency spectrum of the zone plate lens (red data points).

We highlight that the exciton dynamics – the competition between $\gamma_r$ and $\gamma_{nr}$ specifically – directly dictate the excitonic spectral features: $\gamma_r$ controls the amplitude, while the linewidth is dominated by the non-radiative decay rate $\gamma_{nr}$.

Figure 3b displays the squared modulus of the numerically obtained susceptibility of monolayer WS$_2$ (blue solid line), its physically meaningful fit (dashed black line), and the focusing efficiency spectrum superimposed (red data points), showing a very good agreement in their spectral line shape. This analysis confirms that the optical function of the lens is directly governed by the monolayer susceptibility and thereby the excitonic decay rates. The Fano-like line shape in Fig. 3b stems from the superposition of the excitonic and background ($\chi_\infty$) contributions to $\chi$. In contrast to the reflection spectrum in Fig. 3a, where the direct reflection from the substrate and the exciton radiation interfere and result in a near-symmetric reflection peak[26], the directly transmitted light in the focal point is small and can be separated. As such, we emphasize that the focusing efficiency spectrum enables a background-free study of the coherent exciton radiation, whereas the susceptibility extracted from the reflection spectra may also include incoherent exciton radiation.

Our investigation at room temperature demonstrates the unequivocal link between the spectral line shape of the focusing efficiency and the excitonic decay rates. Under these conditions, the non-radiative decay rate for the structured monolayer $WS_2$ (~40 meV) is typically an order of magnitude larger than the radiative decay[7] (~3 meV). This suggests that the optical efficiency of the lens is limited by the exciton's quantum efficiency[9]. While the radiative decay rate is an intrinsic material property, non-radiative decay is dominated by exciton-phonon interactions. Moving towards cryogenic temperatures would thus suppress non-radiative channels for the exciton decay and improve the focusing efficiency spectrum of the lens. To explore this behavior, we use a home-built setup (Fig. 4b) that enables us to systematically study the temperature dependence of the focusing efficiency in transmission, as well as the reflection spectrum, down to sample temperatures of 13 K.

Figure 4a shows the focusing efficiency spectra measured at four roughly equi-spaced temperatures. To provide a representative comparison to the focusing efficiency measurements that contain scattered field contributions from a large area, we measure the temperature-dependent reflectance spectra using a collection area of ~100 μm diameter. Figure 4c shows the temperature-dependent differential reflectance ($\frac{\Delta R}{R_{substrate}} = \frac{R_{WS2} - R_{substrate}}{R_{substrate}}$) obtained by probing a continuous portion of monolayer directly next to the lens (Fig. S6a). Comparison of the differential reflection spectrum at room temperature with Fig. 3a shows a reduced exciton amplitude in the cryostat setup. This is a result of the low numerical aperture of the collection lens in the setup, which collects strong contributions from the rear side of the transparent substrate to the reflection signal (i.e. $R_{substrate}$ is larger).

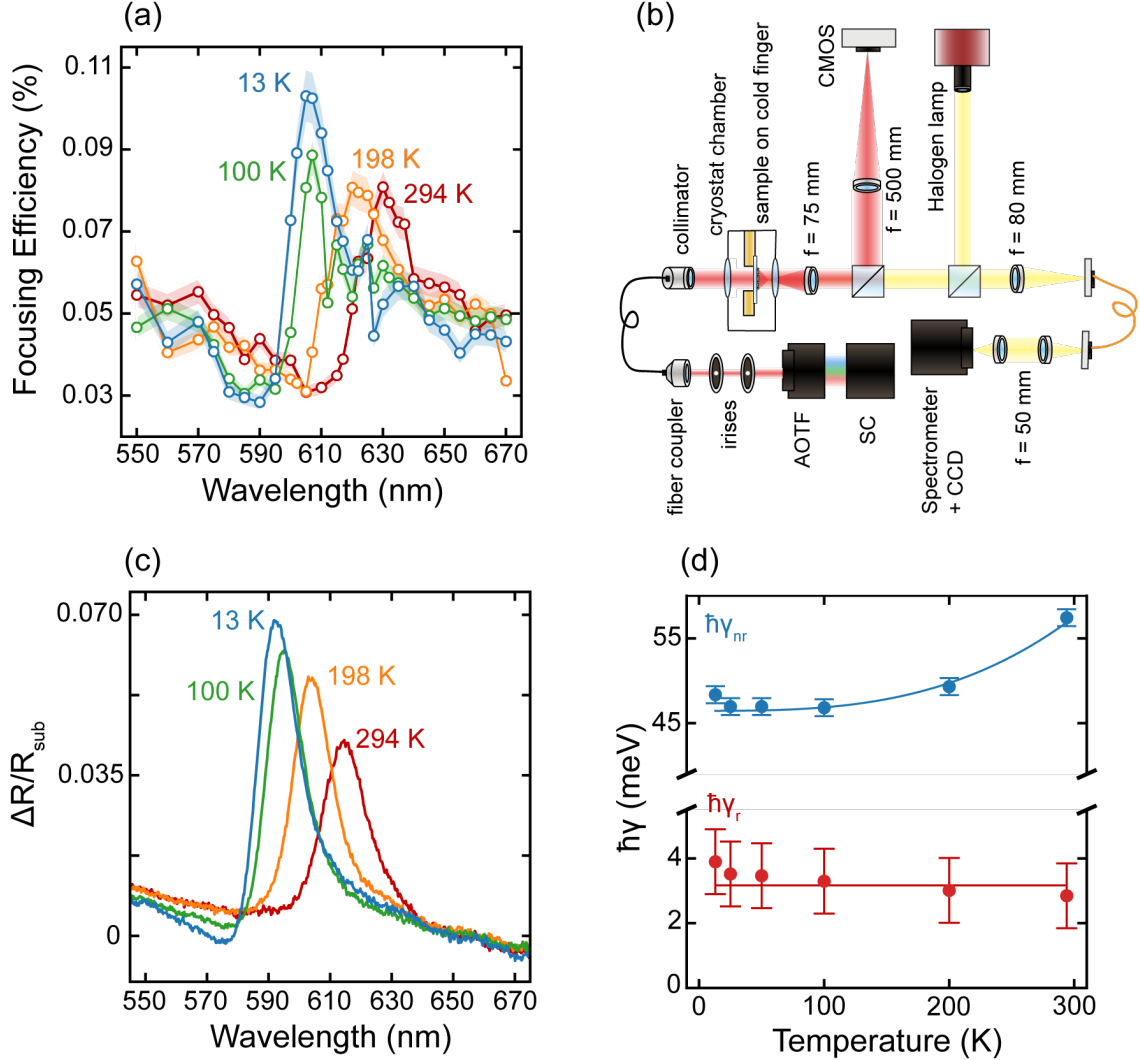

**Figure 4: Temperature dependent measurements.** (a) Focusing efficiency for sample temperatures of 294 K (red), 198 K (orange), 100 K (green), and 13 K (blue). (b) Schematic of the home-built imaging setup used for the temperature- dependent measurements. The focus of the zone plate lens, forming upon monochromatic laser illumination from the rear, is imaged onto a CMOS camera. White light from a halogen lamp reflected by the sample is fiber coupled and directed into a spectrograph. SC: supercontinuum laser; AOTF: acousto-optic tunable filter. (c) Differential reflection spectra for sample temperatures of 294 K (red), 198 K (orange), 100 K (green), and 13 K (blue). (d) Radiative (red) and non-radiative (blue) decay rates (expressed in meV by multiplying with $\hbar$) of $X_A$ for sample temperatures of 294 K, 198 K, 100 K, 50 K, 25 K and 13 K. Solid lines are a guide-to-the-eye.

Comparing Figs. 4a,c, we observe four distinct trends with decreasing temperature. First, we notice a blue shift of the excitonic peak in both the focusing efficiency as well as reflection spectrum. The temperature dependence of the band gap is a recurrent effect in semiconductors and is due to a combination of compressive strain and reduced electron-phonon interactions[27–29]. Second, there is an equally noticeable increase in magnitude. This too originates from the decrease in phonon population, and is corroborated by the temperature dependence of the extracted radiative and non-radiative decay rates (Fig. 4d). The radiative decay rate ($\hbar\gamma_r \sim 3$ meV) is governed by the material's band structure, thus it is independent of temperature. The non-radiative decay rate on the other hand is reduced from $\hbar\gamma_{nr} = 57.4$ meV at room temperature to $\hbar\gamma_{nr} = 48.4$ meV at 13 K due to reduced exciton-phonon scattering. This results in an increased exciton quantum yield and thereby stronger radiation amplitude. Third, an increased asymmetry in the focusing efficiency as well as reflection spectrum is observed for lower temperatures. As the excitonic oscillator strength increases, a stronger contribution of the exciton resonance to the material's susceptibility will result in stronger interference with the non-resonant background scattering. Finally, we observe a concurrent narrowing of the spectral linewidth, corroborated by the rates observed in Fig. 4d. However, the absolute linewidths do not match the narrow linewidths commonly reported for small exfoliated and encapsulated flakes at cryogenic temperatures[7,8] (see discussion below). Despite the limited spectral narrowing, these results clearly demonstrate that the optical function of atomically-thin 2D metasurfaces can be engineered by controlling the excitonic properties.

There are multiple distinct factors that prevent spectral narrowing of the exciton linewidth at lower temperatures. (i) In addition to exciton-phonon interactions, spectral broadening stemming from sample inhomogeneity occurs, which results from substrate effects (e.g. charge transfer, surface roughness, local field fluctuations), and sample deterioration. Consequently, even at lower temperatures, the well-defined excitonic optical transitions can be clouded by broad defect-related and charged-exciton emissions. As shown by Cadiz et al.[22], inhomogeneous contributions can be greatly reduced via encapsulation of the monolayers by hexagonal boron nitride (h-BN). Our large area monolayer, directly deposited on $Al_2O_3$ without encapsulation, reveals linewidths in line with those observed for non-encapsulated TMD monolayers[22], while linewidths down to few meV are only achieved at temperatures of 4 K and in h-BN encapsulated monolayers. (ii) The material quality of large exfoliated flakes tends to exhibit higher defect concentrations than smaller ones.[16] (iii) Sample deterioration that occurred after the room-temperature characterization but before the low-temperature measurements. (iv) In Fig. 4c we probe the temperature-dependent reflectance with a collection diameter of about 100 μm, more than two orders of magnitude larger than our confocal setup (Fig. S2). Via room-temperature reflection mapping spectroscopy of the same area, we find that the spectral variations across such a large collection area add at least 5 meV to the exciton linewidth (Fig. S6).

Thus far, encapsulation has remained rather challenging for large-area monolayers since high-quality hBN flakes are commonly limited to lateral sizes of several 10s of microns. Recent developments in van-der-Waals pick-up methods [30,31] and self-assembled molecular monolayers like 1-dodecanol [32] provide new opportunities to passivate and encapsulate large-area monolayers before nanopatterning. While improved material quality, encapsulation, and absence of background doping[33] can increase the excitonic oscillator strength, optical efficiencies exceeding 10% for atomically-thin elements are prone to be bound to cryogenic temperatures. Alternative and more practical routes towards near-unity efficiencies at room temperature include optical pathlength enhancement by placing the 2D material in a cavity[7,14], or by stacking electrically isolated monolayers[34]. At the same time, we emphasize that even the ~0.1% efficiency elements have important applications in e.g. optical beam tapping[35] and augmented reality[36], where a very small fraction of the signal is redirected or detected[37] while the overall metasurface is optically transparent.

Finally, we note that our model for the exciton decay channels ignores quantum mechanical dephasing as well as the associated momentum-space distribution of the exciton population. The current description assumes a uniform 2D excitonic oscillator[9,38] with only radiative and non-radiative decay that can be described fully classically. While quantum mechanical dephasing of the uniform 2D exciton needs to be accounted for in common reflection experiments where incoherent exciton radiation cannot be removed from the total reflection signal[7,8,39], these dephased photons only provide an incoherent background to the wavefronts that are shaped by the metasurface, and therefore do not contribute to the focusing efficiency. As such, monolayer optical elements provide a unique opportunity to study the coherent exciton light-matter interactions in the absence of incoherent contributions.

In conclusion, we demonstrate how the decay rates of excitons dictate the optical function of large-area and atomically-thin optical elements. Using large-area exfoliation of single crystal $WS_2$, we directly pattern a 500-µm diameter zone plate lens into a monolayer. With temperature-dependent focusing efficiency experiments, we relate the asymmetry in the focusing efficiency spectra to the spectral properties of the susceptibility of the monolayer and retrieve the temperature-dependent exciton decay rates. The optical efficiency increases for lower temperatures, as the exciton exhibits an increased quantum efficiency due to suppressed exciton-phonon scattering. Due to the direct link between exciton resonances and metasurface functionality, large-area, nanopatterned TMDs offer a unique way to access fundamental material properties that, in turn, constitute a new set of design parameters for atomically-thin optical elements.

**Acknowledgements**
This work was funded by institutional funding of the University of Amsterdam. T.B. and J.v.d.G were also supported by a Vidi grant (VI.Vidi.203.027) from the Dutch National Science Foundation (NWO). The Brongersma group was funded by Department of Energy grant DE-FG02-07ER46426 and by a MURI program of the United States Air Force Office of Scientific Research (GrantNo. FA9550-21-1-0312). Preparation of monolayers at Stanford is supported by the Defense Advanced Research Projects Agency (DARPA) under Agreement No. HR00112390108. Part of this work was performed at the Nano@Stanford labs, supported by the National Science Foundation under award ECCS-1542152.